\documentstyle[12pt]{article}

\large
\newcommand{\be}{\begin{eqnarray}}
\newcommand{\ee}{\end{eqnarray}}
\newcommand\del{\partial}

\begin{document}
\setlength{\baselineskip}{21pt}
\pagestyle{empty}
\vfill
\eject
\begin{flushright}
SUNY-NTG-95/4
\end{flushright}

\vskip 2.0cm
\centerline{\Large Effective Lagrangians and Chiral Random Matrix Theory}
\vskip 2.0 cm
\centerline{\bf M.A. Halasz and J.J.M. Verbaarschot}
\vskip .2cm
\centerline{Department of Physics}
\centerline{SUNY, Stony Brook, New York 11794}
\vskip 2cm

\centerline{\bf Abstract}
Recently, sum rules were derived for the inverse eigenvalues of the Dirac
operator. They were obtained in two different ways: i) starting
from the low-energy effective Lagrangian and ii) starting from
a random matrix theory with the symmetries of the Dirac operator.
This suggests that the effective theory can be obtained directly
from the random matrix theory. Previously, this was shown for three
or more colors with fundamental fermions. In this paper we construct the
effective theory from a random matrix theory for two colors in the
fundamental representation and for an arbitrary number of colors in the
adjoint representation.
We construct a fermionic partition function for Majorana
fermions in Euclidean space time. Their reality condition
is formulated in terms of complex conjugation of the second kind.

\vfill
\noindent
\begin{flushleft}
SUNY-NTG-95/4\\
February 1995
\end{flushleft}
\eject
\pagestyle{plain}

\vskip 1.5cm
\noindent
\renewcommand{\theequation}{1.\arabic{equation}}
\setcounter{equation}{0}
\section{Introduction}
\vskip 0.5cm
Although it has been widely accepted that QCD is the correct theory for strong
interactions, the nonlinearities in the interactions have made it very
difficult to obtain accurate results that can be compared to experiment.
In order to obtain rigorous results many researchers
have studied a parameter range
for which the theory simplifies but nevertheless contains the essential
features of QCD. Well-known examples are e.g., the large $N_c$ expansion
\cite{THOOFT-1974a,WITTEN-1979}, the small volume expansion
\cite{LUSCHER,VANBAAL}, 2d theory \cite{THOOFT-1974b}, etc.. Such
methods have provided us with numerous insights in the
physics of the full theory
justifying any new proposal in this direction.

Recently, Leutwyler and Smilga \cite{LS} proposed
to focus on the quark mass dependence
of the Euclidean QCD partition function
close the the chiral limit (quark masses $m \ll \Lambda_{\rm QCD}$)
in volumes with length scale $L$ much
larger than a typical hadronic length scale ($\Lambda^{-1}_{\rm QCD}$)
but still much smaller than the pion Compton wave length ($\sim
1/\sqrt{m \Lambda_{QCD}}$).
The first condition assures that only the low-lying excitations,
in particular the Goldstone modes, contribute to the partition function,
whereas the second condition allows us to ignore the kinetic terms in
the Lagrangian. The mass dependence of the partition function
is then given by the Lagrangian \cite{LS}
\be
{\cal L}= m V \Sigma {\rm Re}({\rm Tr U}),
\ee
where $\Sigma$ is the vacuum expectation value of $\bar \psi \psi$
which is assumed to be nonzero,
and the unitary matrix $U$ parameterizes the Goldstone fields.
The question that was asked in \cite{LS} is to what extent the knowledge of the
finite volume partition function puts constraints of the spectrum of the
Euclidean Dirac operator. By expanding both the QCD partition function
and the partition function corresponding to (1.1) in powers of $m$,
it was found,
by equating the coefficients, that the inverse powers of the eigenvalues
satisfy the Leutwyler-Smilga sum rules.
Because of the Banks-Casher formula \cite{BANKS-CASHER}  the smallest
nonzero eigenvalue is of order $1/\Lambda_{QCD}^3 L^4$, whereas in the absence
of interactions, the smallest eigenvalue is of order $1/L$. The sum rules
hold for eigenvalues well below the latter scale, which for sufficiently large
volumes is well separated from the scale of the smallest eigenvalues.

In previous works \cite{SVV,VZ,V,SV}
we have investigated several questions related to the
nature of these sum rules.
In particular, it was argued that since the
effective theory is solely based on the symmetries of QCD, the sum rules
depend on the $symmetries$ of the Dirac operator as well.
So when one starts out with a
theory with the same global symmetries as QCD but no other dynamical input
one should arrive at exactly the same sum rules.
We have constructed such theories: chiral random matrix theories,
which apart from the chiral symmetry and possible anti-unitary symmetries
also contain a remnant of the topological structure of QCD.
In the framework of random matrix
theory one has three different universality classes,
those with real, complex or quaternion real matrix elements \cite{DYSON-three}.
For $SU(2)$ with fundamental fermions the matrix
elements are real, they are complex for
more than two colors with fundamental fermions,
whereas they are quaternion real for adjoint fermions.
Indeed, it was shown that the sum rules that are obtained from the
chiral random matrix theories \cite{V}
coincide with those obtained from the effective theory \cite{LS,SV}.
It should be noted that the sum rules for $SU(2)$ were found only
after the introduction of chiral random matrix theories \cite{V}.
The three classes of random matrix
theories correspond to the three different schemes
of chiral symmetry breaking, which were discussed before
in the literature \cite{Shifman-three}.

This suggests that it is possible to derive the finite volume
partition function directly
from the random matrix theory. In fact, this task has been performed
for QCD with three or more colors \cite{SVV}.
However, for QCD with two colors
or QCD with adjoint fermions, the situation is more complicated, and up to
now such equivalence has not been proved. The main objective of this
paper is to show that also in these two cases there is a one to one
correspondence between chiral random matrix theory and the low energy
finite volume partition function.

The main complication is the presence of anti-unitary symmetries in
the QCD Lagrangian. For $N_c =2 $ this symmetry leads to a real Dirac
operator and apart from a somewhat more complicated algebra it is
straightforward to obtain the effective Lagrangian. To deal with
adjoint fermions we have to face the well-known assertion that Majorana
fermions do not exist in Euclidean space time \cite{Ramond}.
However, this statement refers
to the transformation properties of a Dirac spinor under Lorentz
transformations. It does not exclude the possibility to
write down a partition function for Majorana fermions
in terms of Grassmann integrals (see, for example, \cite{VaZ}).
Starting from the observation that the adjoint
Euclidean Dirac operator is anti-symmetric up to a charge conjugation
matrix, we indeed succeeded to do this.
In fact, using conjugation of the
second kind \cite{Berezin,Efetov,VWZ},
the Majorana condition in Euclidean space time is completely
analogous to the one in Minkowski space time.

In a previous work \cite{SV},
it was observed that the simplest sum rule for each of the
three cases could be summarized by one formula involving the dimension
of the Goldstone manifold of the theory.
In order to show that this was no coincidence,
we present a derivation that leads to this result naturally.

The structure of this paper is as follows.
The symmetries of the Dirac operator for fundamental and adjoint fermions
are discussed in sections 2 and 3, respectively. In section 4 we discuss the
random matrix theory with these symmetries as input. The effective theories
for the three different cases are derived in sections 5a, 5b and 5c.
A general derivation of the simplest sum rule is given in section 6 and
concluding remarks are made in section 7.

\vskip 1.5cm
\noindent
\renewcommand{\theequation}{2.\arabic{equation}}
\setcounter{equation}{0}
\section{Symmetries of the Dirac operator for fundamental fermions}
\vskip 0.5 cm

In this section we study the symmetries of the Euclidean
Dirac operator\footnote{Our conventions are
that the Euclidean gamma matrices are
Hermitean and satisfy the anti-commutation relations $\{\gamma_\mu,
\gamma_\nu\} = 2\delta_{\mu\nu}$. We use a chiral representation in which
$\gamma_5 $ is diagonal.}

\be
i\gamma_{\mu}D_{\mu} =
i \gamma_{\mu} \del_\mu+ \gamma_\mu A_\mu,
\ee
for a fixed background field $A_\mu$ in the fundamental representation of
$SU(N_c)$. For $N_c \ge 3$, the $N_c \times N_c$ matrix
$A_\mu$ is complex valued, and  the only symmetry of the Dirac operator
is the chiral symmetry
\be
\{i\gamma D, \gamma_5\} = 0.
\ee
Because of this the eigenvalues occur in pairs $\pm \lambda$ or are zero. The
eigenfunctions are related by $\phi_\lambda = \gamma_5 \phi_{-\lambda}$. For
$\lambda = 0$ we have the possibility that $\gamma_5 \phi_{\lambda=0} =
\pm \phi_{\lambda=0}$, resulting in a fermionic
zero mode that is either right handed or left handed. It is well known that
this happens in the field of an instanton \cite{THOOFT-1976}.

Because of the symmetry (2.2), the matrix representation of the Dirac operator
simplifies in a chiral basis $\{\phi_{R\, k},\, \phi_{L\, k}, \}$
with $\gamma_5 \phi_{R\, k} = \phi_{R\, k}$ and
$\gamma_5 \phi_{L\, k}  = -\phi_{L\, k}$. The action becomes
\be
\sum_{f,g=1}^{N_f}
\sum_{kl}\left (\begin{array}{c}   \chi_R^f \\ \chi_L^f \end{array}\right )^*_k
\left (\begin{array}{cc} i{\bf m}_{fg}^* & D_{RL} \\ D_{LR} &
i{\bf m}_{fg} \end{array}
\right )_{kl}
\left (\begin{array}{c}   \chi_R^g \\ \chi_L^g \end{array}\right )_l,
\label{funpart}
\ee
where we have also included a mass matrix
${\bf m}^*(1+\gamma_5)/2 + {\bf m}(1-\gamma_5)/2$.

The matrix elements of $D_{RL}$ given by
\be
\int d^4 x\phi_{R\,k}^{*}\, i\gamma D \,\phi_{L\, l}
\ee
are flavor diagonal.
For a Hermitean Dirac operator we have $D_{LR} = D_{RL}^\dagger$.
In the presence of fermionic zero modes the number of
left handed and right handed states is not necessary equal.
In general, the matrix $D_{RL}$ is a rectangular
one. Indeed, the total number of zero eigenvalues of the matrix in
(2.3) in the chiral limit $(m\rightarrow 0)$
is given by the absolute value of the difference in dimensionality
of the left handed and the right handed Hilbert spaces.

For $N_c =2$ the Euclidean Dirac operator is given by
\be
i\gamma_\mu D_\mu  = i \gamma_{\mu} \del_\mu+ \gamma_\mu A_\mu^a
\frac{\tau_a}2,
\ee
where  $\tau_k$ are the Pauli spin matrices ($\tau_1 \tau_2 = i \tau_3$,
etc.). In addition to the chiral symmetry (2.2),
this Dirac operator possesses the anti-unitary symmetry:
\be
[i\gamma D, C \tau_2 K] =0 \quad {\rm for } \quad N_c = 2.
\ee
Here, $C$ is the charge conjugation matrix ($C = \gamma_2 \gamma_4$
and $C^2 = -1$) and $K$
is the complex conjugation operator. The anti-unitary operator $C\tau_2 K$
satisfies
\be
(C\tau_2 K)^2 = 1
\ee
{}From the analogy with the time reversal symmetry in quantum mechanics
\cite{Porter} it is clear that this condition allows us to choose
a basis in which $i\gamma D$ is real.
To show this, we construct a basis $\{\psi_k\}$ that diagonalizes $C\tau_2 K$
with eigenvalues one,
\be
C\tau_2 \psi_k^* = \psi_k.
\label{cpsi}
\ee
We start with an arbitrary basis vector $\phi_1$, then $\psi_1 =
\phi_1+C\tau_2\phi_1^*$ satisfies (\ref{cpsi}). Next we choose $\phi_2$
perpendicular to $\psi_1$. The second basis vector satisfying (\ref{cpsi}) is
given by $\psi_2 = \phi_2+C\tau_2\phi_2^*$, and it can be shown easily that
it is orthogonal to $\psi_1$. The next basis vector is obtained from
$\phi_3$ perpendicular to both $\psi_1$ and $\psi_2$, etc..

By using (2.6) in the form
\be
\tau_2 C \,i\gamma D C \,\tau_2 = - (i\gamma D)^*,
\ee
for which no analogy exists for $N_c \ge 3$,
it follows immediately that the matrix elements (2.4) of the Dirac operator
are real in this basis. Note that, because $[\gamma_5, C\tau_2 K] = 0$,
the above construction holds true for a chiral basis.
For $N_c = 2$, the fermionic part of the action is therefore
given by (2.3), but with $real$ valued matrices $D_{RL}$ and $D_{LR}$.
This makes it possible to rewrite (2.3) for ${\bf m =0}$ as
\be
\sum_{f=1}^{N_f}
\sum_{kl}\left (\begin{array}{c}   \chi_R^f \\ \chi_R^{f\,*}
\end{array}\right )_k D_{RL}^{kl}
\left (\begin{array}{c}   -\chi_L^{f\,*} \\ \chi_L^f \end{array}\right )_l,
\ee
showing that the chiral symmetry group is enlarged to $U(2N_f)$. A mass matrix
proportional to the identity breaks this symmetry to $Sp(2N_f)$. Below
we will show that the same breaking pattern occurs by the formation of
a quark condensate.
It is important that the construction of the above basis
relies on the $symmetries$ of the Dirac operator only, and that
the Dirac matrix becomes real for an arbitrary $SU(2)$ color gauge field.

\vskip 1.5cm
\renewcommand{\theequation}{3.\arabic{equation}}
\setcounter{equation}{0}
\section{Symmetries of the Dirac operator for Majorana fermions}
\vskip 0.5 cm
The situation with Majorana fermions in the adjoint representation
of the gauge group is much more complicated. The
reality condition  that defines Majorana fermions and their partition function
in Minkowski space
cannot be simply generalized to Euclidean space time, and
in the literature one can find statements that Majorana fermions do
not exist in Euclidean space \cite{Ramond}.
Below we will make this more explicit and provide a construction
of a fermionic partition function that works as well in Euclidean space
as in Minkowski space.

The Euclidean Dirac operator in the adjoint representation
given by
\be
i\gamma_\mu D_\mu = i \gamma_\mu( \del_\mu \delta_{ab} + f_{abc} A_\mu^c),
\label{addirac}
\ee
where $A_\mu^c$ is an $SU(N_c)$ background field distributed
according to the gluonic action. The essential difference from the Dirac
operator in the fundamental representation is that the long derivative
is anti-symmetric under transposition, which is also true in Minkowski space.
Below we will exploit this by
using that the full Dirac operator is anti-symmetric up to a constant matrix
with determinant one.

Apart from the chiral symmetry,
\be
\{i\gamma_\mu D_\mu, \gamma_5\} = 0,
\ee
the Dirac operator (\ref{addirac}) also has an anti-unitary symmetry
\be
[i\gamma D, CK] = 0.
\label{adsym}
\ee
First notice that, because
\be
(CK)^2 = -1,
\label{ck}
\ee
it is not possible to repeat the construction
for $N_c =2$ in the fundamental
representation resulting in a real matrix.
To see this, remember that the construction was based on the
diagonalization of the anti-unitary symmetry operator. Now, let us
assume that we can find an eigenvector $CK \phi = \lambda \phi$. Then,
\be
(CK)^2 \phi = CK \lambda \phi = \lambda^* \lambda \phi,
\ee
which in view of (\ref{ck}) leads to an obvious contradiction. As a corollary
it follows that $\phi$ and $CK\phi$ are linearly independent.

To analyze the implications of the symmetry (\ref{adsym}) in Euclidean
space, we first  address the issue of obtaining a fermionic action
for Majorana fermions in Minkowski space.
In this case the Dirac operator in the adjoint representation satisfies the
commutation relation
\be
[i\gamma_\mu D_\mu, \gamma_2 K] =0,
\ee
and the anti-unitary charge conjugation operator
\be
(\gamma_2 K)^2 =1.
\label{gk}
\ee
Therefore, the operator $\gamma_2 K$ can be diagonalized, and
a superselection rule can be imposed that restricts the partition function
to states with eigenvalue 1. Such states, called Majorana fermions,
can be parameterized by
\be
\psi^M = \left ( \begin{array}{c} \chi_R \\ -\sigma_2 \chi^*_R \end{array}
\right ).
\label{basism}
\ee
Since $\gamma_2 K$ also commutes with Lorentz transformations\footnote{In a
chiral basis they are given by \cite{Ramond} $\psi_L \rightarrow \Lambda_L
\psi_L$ and $\psi_R \rightarrow \Lambda_R\psi_R$ with
$\Lambda_L =\exp(\frac i2\vec \sigma (\vec \omega -i\vec\nu))$ and
$\Lambda_R =\sigma_2 \Lambda_L^*\sigma_2$.},
 $\psi^M$ transforms as a Dirac spinor under the Lorentz group.
This result is usually stated  as
follows \cite{Ramond,Ramond-1994}:
$\chi_L$ and $-\sigma_2 \chi_R^*$ transform in the same way
under Lorentz transformations.
The Majorana Lagrangian is thus given by
\be
\bar \psi^M i\gamma D \psi^M,
\ee
where $\bar \psi^M = \gamma_4\psi^{M\, *}$.
Consistent with (\ref{gk}), the Minkowski Dirac operator is real in the
basis (\ref{basism}).

Let us pursue the construction of Majorana fermions and their action
along a different route. The crucial
observation is that the matrix $\gamma_2\gamma_4 i\gamma D$ is
antisymmetric under transposition.
This allows us to write down a fermionic action with
half as many degrees of freedom
\be
{\det}^{1/2}i\gamma D = {\det}^{1/2}\gamma_2\gamma_4i\gamma D =
\int {\cal D} \psi \exp[ \psi \gamma_2\gamma_4 i\gamma D \psi].
\label{majmin}
\ee
Lorentz invariance follows immediately form the fact that
that $\gamma_2\gamma_4\psi$ and $\gamma_4\psi^*$
transform in the same way under Lorentz transformations.
This can be made more explicit by introducing a conjugation operation such that
\be
\gamma_4 \psi^* = \gamma_2\gamma_4\psi,
\ee
which, of course, is precisely the Majorana condition.
Before attacking the problem of 4 dimensional Euclidean Majorana fermions
let us first consider the problem in 1+1 dimensional Minkowski space
and two dimensional Euclidean space.

In 1+1 dimensional Minkowski space time, the Dirac operator
in the adjoint representation, $i\gamma_\mu D_\mu$
with $\gamma-$matrices defined by
$\{\gamma_\mu, \gamma_\nu\} = 2 g_{\mu\nu}$ (We use the convention
$\gamma_0 =  \sigma_2, \,\, \gamma_1 = i\sigma_1$)
has the anti-unitary symmetry
\be
[i\gamma D, K] = 0.
\ee
Since $K$ also commutes with Lorentz transformations\footnote{For our
representation of the gamma matrices, Lorentz transformations of
spinors are given by $\psi \rightarrow\exp( -\omega \sigma_3) \psi$.}
the Lagrangian for
Majorana fermions, defined by the condition $K\psi^M =\psi^M$, is given by
\cite {Smilga}
\be
{\cal L}^M = \bar \psi^M i\gamma D \psi^M.
\ee
Here, $\bar \psi^M = \gamma_0 \psi^{M\,*} = \gamma_0 \psi^M$.

As in 3+1 dimensions, we could have followed an alternative
route leading to the same Majorana action. The starting point is the
observation that $i\sigma_2 i\gamma D$ is
antisymmetric with respect to transposition
which allows us to write the square root of the fermion
determinant as a Grassmann integral with only half as many degrees
of freedom:
\be
{\det}^{1/2}i\gamma D=  {\det}^{1/2}i\sigma_2 i\gamma D = \int {\cal D} \psi
\exp[ \psi i\sigma_2 i\gamma D \psi].
\ee
This expression is Lorentz invariant. Indeed, $\psi$ satisfies
the Majorana condition.

Let us now proceed to Euclidean Majorana fermions in two dimensions.
The gamma matrices are defined by
$\{\gamma_\mu, \gamma_\nu\} = 2\delta_{\mu\nu}$, and we use the representation
$\gamma_0 = \sigma_2, \,\, \gamma_1 = \sigma_1$. The Dirac operator
satisfies the commutation relation
\be
[i\gamma_\mu D_\mu, i\sigma_2 K] = 0.
\label{3.15}
\ee
Because $(i\sigma_2 K)^2 = -1$ we cannot impose a Majorana condition of the
form $i\sigma_2 K \psi = \psi$. However, the matrix $i\sigma_2 i\gamma D$
is anti-symmetric
with allows us to halve the number of fermionic degrees of freedom
in the partition function
\be
{\det}^{1/2}i\gamma D = {\det}^{1/2}i\sigma_2 i\gamma D = \int {\cal D} \psi
\exp[ \psi i\sigma_2 i\gamma D \psi].
\ee
This action is invariant under
Euclidean Lorentz transformation, i.e. under $\psi \rightarrow
\exp(i\sigma_3 \phi) \psi$.
This can be made more explicit
by introducing a conjugation operator such that
\be
\psi^* = -i\sigma_2\psi.
\label{star2d}
\ee
In components, the equations read
\be
\psi_1^* = -\psi_2,\qquad
\psi_2^* = \psi_1,
\ee
which forces us to impose the consistency condition $\psi_k^{**} = - \psi_k$.
This conjugation, called conjugation of the second kind, is well-known
in the mathematical literature on Grassmann variables
\cite{Berezin}, and has been used extensively in the supersymmetric
formulation of random matrix theories \cite{Efetov, VWZ}.
The Majorana constraint (\ref{star2d})
\be
i\sigma_2 K \left (\begin{array}{c} \psi_1 \\ \psi^*_2 \end{array} \right )
= \left (\begin{array}{c} \psi_1 \\ \psi^*_2 \end{array} \right ),
\ee
makes the symmetry (\ref{3.15}) manifest at the level of the partition
function.

Let us now return to 4 Euclidean dimensions. The strategy should be clear:
we construct an anti-symmetric operator that differs from the Euclidean
Dirac operator only by a factor with unit determinant. As can be seen
from the following theorem, it is not an accident that this works.

\noindent
{\it Theorem}. Consider a Hermitean operator $H$ such that
$[H, AK]= 0$ and $A^\dagger A = A A^\dagger = 1$. If $(AK)^2 = -K^2$ then
$(HA)^T = - HA$. If, moreover $A$ or $iA$ are orthogonal then also
$(AH)^T = - AH$.

The proof of this theorem is immediate. It can be applied to the
Dirac operator in Euclidean space time. However, since going from Euclidean
to Minkowski space just amounts to multiplying the  spacial gamma matrices
by a factor $i$, in both cases the Dirac operator behaves the same under
transposition.

Using this theorem, one concludes from (3.3) that
$C i\gamma D$ is an antisymmetric matrix, which allows us
to write down a fermionic partition function with only half as
many degrees of freedom
\be
{\det}^{1/2}i\gamma D= {\det}^{1/2}C i\gamma D = \int {\cal D} \psi
\exp[ \psi C i\gamma D \psi].
\label{majpart}
\ee
This construction also works in the presence of a mass term.
It is straightforward to verify that (3.20) is invariant under Euclidean
Lorentz transformations. As before, this can be made more explicit by
choosing the components of $\psi$ such
that they satisfy the conjugation equation
\be
C\psi^* = \pm\psi.
\label{majcon}
\ee
A solution with eigenvalue $+1$ can be parametrized by
\be
\psi = \left ( \begin{array}{c} \psi_1 \\ \psi_1^* \\ \psi_2 \\ -\psi_2^*
\end{array} \right),
\ee
which can be viewed as a Majorana constraint
on the fermionic integration variables. As was the case
for Euclidean fermions in 2 dimensions, we have to impose
conjugation of the $second$ kind on the Grassmann variables (otherwise the
equation $CK\psi = \pm \psi$ does not have solutions, see eq. (3.5)).

To display the matrix structure of the Dirac operator we express the
field in the Lagrangian (\ref{majpart}) in terms of a complete
set with Grassmannian
coefficients.
An expansion consistent with the Majorana constraint
(\ref{majcon}) is given by (because $(CK)^2 = -1$, the c-number
functions $\phi$ and $ C\phi^*$ are linearly independent)

\be
\psi_R = \sum_k \phi_{R\,k} \chi_{R\,k} + C \phi_{R\,k}^* \chi_{R\,k}^*
\ee
for right handed fermions and,
\be
\psi_L = \sum_k \phi_{L\,k} \chi_{L\,k} + C \phi_{L\,k}^* \chi_{L\,k}^*
\ee
for left handed fermions. In this basis the fermionic piece of the action
is given by
\be
\left ( \begin{array}{c} \chi_R^f \\ \chi_R^{f\,*} \\ \chi_L^f \\ \chi_L^{f\,*}
\end{array} \right )^*_k
\left ( \begin{array}{cc} {\bf m}^*_{fg}
& D_{RL} \\ D_{LR} & {\bf m}_{fg} \end{array} \right )_{kl}
\left ( \begin{array}{c} \chi_R^g \\ \chi_R^{g\,*} \\ \chi_L^g \\ \chi_L^{g\,*}
\end{array} \right )_l,
\label{majdir}
\ee
where each $2\times 2$ block of $D_{RL} $ is given by
\be
(D_{RL})_{kl} = \left ( \begin{array}{cc}
\int \phi_{R\,k}^*\, i\gamma D \, \phi_{L\,l} & \int \phi_{R\,k}^*\,
i\gamma D \, C\phi_{L\,l}^* \\
\int \phi_{R\,k} \, C^T i\gamma D \, \phi_{L\,l} &\int \phi_{R\,k} \,
C^T i\gamma D \, C\phi_{L\,l}^*
\end{array} \right ),
\ee
and $D_{LR} = D_{RL}^\dagger$. In (\ref{majdir}) we have included the flavor
indices. Using that
\be
C^T i\gamma D C = (i\gamma D)^*, \\
C^T i\gamma D = -(i\gamma D)^* C,
\ee
we find that each $2\times 2$ block in $ D_{RL}$ is quaternion real, i.e. of
the form $a_0 + ia_k \sigma_k$ with $a_\mu$ real. The mass matrix is
proportional to the unit quaternion but is generally not diagonal
in the flavor indices. Since
the product of Grassmann variables that multiplies the mass matrix is symmetric
in the flavor indices, the mass matrix can be taken symmetric as well.

Because the Grassmannian vectors on the left and on the right in (3.25)
do not transform independently
under transformations in flavor space, the chiral symmetry for ${\bf m}=0$
is reduced to $U(N_f)$.  For a nonzero mass matrix that is
a multiple of the identity,
only an invariance under the $O(N_f)$ subgroup remains. Below we will see
that the same breaking is achieved by the formation of a nonzero chiral
condensate.

The eigenvalues of the Dirac operator (\ref{majdir}) are doubly degenerate
and, for zero mass,
occur in pairs $\pm \lambda$ or are zero. The number of zero eigenvalues
is equal to $twice$ the absolute value of the difference of the dimensionality
of the right handed and the left handed space.

\vskip 1.5 cm
\renewcommand{\theequation}{4.\arabic{equation}}
\setcounter{equation}{0}
\section{Chiral random matrix theories}
\vskip 0.5 cm
The matrix elements of the Dirac operator fluctuate over the ensemble
of gauge fields. In this section we introduce a
model partition function with the $symmetries$ of the Dirac operator
but with matrix elements fluctuating according to independent gaussian
distributions. As discussed in the introduction,
we expect that certain low energy quantities,
such as the fluctuations of the eigenvalues on a microscopic scale,
do not depend on the dynamics of QCD interaction and can be calculated in
this simplified model.

The QCD partition function can be written as
\be
Z_{QCD} = \sum_{\bar \nu} e^{i\bar\nu\theta} Z_{QCD}(\bar\nu),
\ee
where the partition function in the sector with $\bar\nu$ fermionic zero modes
is defined by
\be
Z_{QCD}(\bar\nu)= \langle \prod_{f=1}^{N_f} m_f^{\bar\nu} \prod_{\lambda_n > 0}
(\lambda_n^2 +|m_f|^2)\rangle_A.
\ee
The eigenvalues of the mass matrix are denoted by $m_f$, and
the average goes over all gauge fields with topological charge $\nu$
weighted according to the gluonic action. For fundamental fermions
$\bar\nu = \nu$, but
for adjoint fermions, with
each of the doubly degenerate eigenvalues included only once in the fermion
determinant, we have $\bar \nu = N_c\nu$ (see \cite{LS}).

In sectors with a nonzero total topological charge, the
number of right handed modes in (\ref{funpart}) and
(\ref{majdir}) differs from the number of
left handed modes. The random matrix partition function with
$n$ and $n+\bar\nu$ of such modes, respectively, and
$N_f$ flavors is defined by
\be
Z_\beta(\bar\nu, N_f) = \int {\cal D} T \prod_{f=1}^{N_f}{\det} \left (
\begin{array}{cc} m_f^* &i T \\i T^\dagger & m_f \end{array} \right )
\exp[-\frac{n\beta\Sigma^2}2 {\rm Tr} T T^\dagger],
\label{zrandom}
\ee
where $T$ is an $n\times (n+\bar\nu)$ matrix. The integration over $T$ is
according to the Haar measure.  The matrix elements of $T$ are real for $\beta
=1$ corresponding to
QCD with $N_c = 2$ in the fundamental representation.
They are complex for $N_c \ge 3$ in the fundamental
representation ($\beta = 2$). For gauge fields in the adjoint
representation the matrix elements of the Dirac operator are
quaternion real (see (3.27)), and the matrix
elements $T_{ij}$ in (\ref{zrandom}) are chosen
quaternion real as well ($\beta = 4$).
If we write the quaternions in terms
of $2\times 2$ matrices, the matrix $T$ is a $2n\times 2(n+\bar \nu)$
matrix. Of course, the masses $m_f$ are multiplied by
the quaternion unit matrix.
The determinant in (\ref{zrandom})
for $\beta = 4$ is the so called
${\rm Qdet}$ and the trace is the ${\rm QTr}$.
It can be shown that for a quaternion real matrix $A$ that
${\rm Qdet}^2 A = \det A$ \cite{Dyson-1970}. In a $2\times 2$ matrix
representation of the quaternions the ${\rm QTr}$ is just one half
the ordinary trace.
Note that $C$ times the unit in the flavor indices multiplied by
the matrix in (3.26) is antisymmetric
which makes the square root of its determinant is well defined.

In analogy with the classical
random matrix ensembles, these ensembles
will be called, the chiral orthogonal ensemble
(chGOE), the chiral unitary ensemble (chGUE) and the chiral symplectic
ensemble (chGSE), for $\beta = 1,\, 2$, and 4, respectively.

We will identify the total number of modes, $2n$, (we always have
$\bar \nu \ll n$) with the volume of space time. This corresponds to
choosing units in which the density of the low-lying modes is equal to one.
Below we will see that the parameter $\Sigma$ can be identified
as the chiral condensate.

The matrix ensembles described by the partition function (\ref{zrandom})
are equivalent to what is known in the random matrix
literature as the Laguerre ensembles.
They first were introduced by Fox and Kahn \cite{FOX-KAHN-1964}. The simplest
case $\beta = 2$ was analyzed in \cite{BRONK}. An analysis of the $\beta =1$
case in the context of the microscopic spectral density of the Dirac operator
was given in \cite{V}.
An analysis which also includes many other correlation functions
for $\beta =1$ and $\beta= 4$ was performed in a series
of papers by Nagao and coworkers and Forrester \cite{NAGAO, Forrester}.
Other results for the chiral random matrix ensembles
have been obtained in terms of a supersymmetric
formulation \cite{andreev}
The recent work on this subject
is based on results by Mehta and Mahoux \cite{Mahoux-Mehta-1991} who
introduced the skew orthogonal polynomials originally invented by Dyson
\cite{DYSON-skew}.

The fermion determinant in (4.1) can be written as a Grassmann integral
\be
Z_\beta(\bar\nu, N_f) =
\int {\cal D} T {\cal D} \psi^* {\cal D} \psi
\exp[- i\sum_{f,g=1}^{N_f}\psi^{f\,*}_i \left (
\begin{array}{cc} i{\bf m}^*_{fg} & T \\ T^\dagger & i{\bf m}_{fg} \end{array}
\right )_{ik}\psi^g_k
-\frac{n\beta\Sigma^2}2 {\rm Tr} T T^\dagger].
\label{ranpart}
\ee
For adjoint fermions ($\beta =4$) the matrix elements of $T$ are quaternion
real, and  each component $\psi^f_i$ is a vector of length
four given by (see (3.26))
\be
\psi_i^f = \left ( \begin{array}{c} \chi^{f}_{R\,i} \\ \chi^{*\,f}_{R\,i}
\\ \chi^{f}_{L\,i} \\ \chi^{*\,f}_{L\,i} \end{array}
\right ).
\ee
They satisfy a reality condition similar to the Majorana constraint.
In order to assure ourselves of a positive definite fermion determinant
for $\nu = 0$ we have included a factor $i$ in front of the fermionic
action.
For $\beta = 1$ and $\beta = 2$ the components of $\psi$ and $\psi^*$ are
independent integration variables.

\vskip 1.5 cm
\renewcommand{\theequation}{5.\arabic{equation}}
\setcounter{equation}{0}
\section{Effective theory}
\vskip 0.5 cm
In this section we derive the effective theory corresponding to the
random matrix partition function (\ref{ranpart}).
We proceed by averaging over the
matrix $T$ resulting in a four-fermion interaction which can be
made gaussian at the expense of a new bosonic integration variable.
After performing the Grassmann integrals the resulting
theory is amenable to a saddle point approximation in which the integrals
over the soft modes are kept and the integrals over the hard modes are
done to gaussian order.

For completeness we start the discussion with the
simplest case $\beta = 2$ (section 5a)
which was already analyzed in ref. \cite{V}.
In sections 5b and 5c we discuss the cases $\beta = 1$ and $\beta = 4$,
respectively.

\vskip 1.0 cm
\subsection{Effective theory for $\beta = 2$}
\vskip 0.5 cm
After averaging over the matrix elements of the Dirac operator the partition
function becomes
\be
Z_2(\bar\nu, N_f) \sim
\int {\cal D} \psi^* {\cal D}\psi \exp[- \frac 2{n\Sigma^2 \beta}
\psi^{f\,*}_{L\, k}\psi^{f}_{R\, i}
\psi^{g\,*}_{R\, i}\psi^{g}_{L\,k }+({\bf m}^*_{fg}\psi^{f\,*}_{R\, i}
\psi^{g}_{R\, i}+{\bf m}_{fg}\psi^{f\,*}_{L\, k}\psi^{g}_{L\, k})].
\nonumber \\
\ee
Here and below we have used the $\sim$ sign
in order to indicate that constant factors have been absorbed in the
normalization of the partition function.
The term of fourth order in the Grassmann variables can be rewritten as the
difference of two squares,
\be
\psi^{f\,*}_{L\, k}\psi^{g}_{L\, k}
\psi^{g\,*}_{R\, i}\psi^{f}_{R\,i } =
\frac 14(\psi^{f\,*}_{L\, k}\psi^{g}_{L\, k}
+\psi^{g\,*}_{R\, i}\psi^{f}_{R\,i })^2
-\frac 14(\psi^{f\,*}_{L\, k}\psi^{g}_{L\, k}
-\psi^{g\,*}_{R\, i}\psi^{f}_{R\,i })^2
\label{split}
\ee
Using the Hubbard-Stratonovitch transformation, each of the squares can
be linearized by introducing an additional Gaussian
integral
\be
\exp(-AQ^2) \sim \int d\sigma\exp(-\frac{\sigma^2}{4A} - iQ \sigma).
\label{Hubbard}
\ee
This results in the partition function
\be
Z_2(\bar\nu, N_f) &\sim&
\int {\cal D} \sigma_1{\cal D} \sigma_2
{\cal D} \psi {\cal D} \psi^* \exp[ -\frac{n\Sigma^2\beta}2 {\rm Tr} (\sigma_1
\sigma_1^T+\sigma_2 \sigma_2^T) \nonumber\\ &+&\psi^{f\,*}_{R\,i}
\psi^{g}_{R\,i}(\sigma_1^{fg} - i\sigma_2^{fg} +{\bf m}^*_{fg})
+\psi^{f\,*}_{L\,k}
\psi^{g}_{L\,k}(\sigma_1^{fg} + i\sigma_2^{fg} +{\bf m}_{fg})],
\ee
where $\sigma_1$ and $\sigma_2$ are arbitrary real $N_f\times N_f$ matrices.
The integration over the Grassmann variables yields
\be
Z_2(\bar\nu, N_f) \sim \int {\cal D} A {\det}^{n+\nu}
(A^\dagger + {\bf m}^*   ){\det}^{n}(A +{\bf m})
\exp [-\frac{n\Sigma^2\beta}2 {\rm Tr}A A^\dagger],
\ee
where $A$ is an arbitrary complex matrix. It can be diagonalized
according to
\be
A= U \Lambda V^{-1}.
\ee
In order to have the same number of degrees of freedom on both sides of
the equation we chose $U \in U(N_f)$ and $V\in U(N_f)/(U(1))^{N_f}$.
All matrix elements of the diagonal matrix $\Lambda$ are real non-negative.
For $n\rightarrow \infty$ and $\|{\bf m}\|\Sigma \ll 1$,
the integral over the eigenvalues can be performed  by a saddle point
approximation at ${\bf m}=0$.
The solutions of the saddle point equation for $\Lambda$
are given by
\be
\Lambda_k = \pm \frac 1{\Sigma},
\ee
but only the solution with all signs positive is inside the integration
manifold. At this point,  the  integral only depends
on the combination $UV^{-1}$
which allows us to absorb $V$ in the integration over $U$ and perform the
$V$ integration. For small masses, $m_f\Sigma \ll 1$ (the eigenvalues of
the mass matrix are denoted by $m_f$),
the integral over $U$ is soft and cannot be done by a saddle point method.
However, in this limit we can
expand the determinant to first order in ${\bf m}$ which leads to the
partition function
\be
Z_2(\bar\nu, N_f) \sim \int d \theta \exp(i\bar\nu\theta)\int_{U\in SU(N_f)}
{\cal D} U \exp [n \Sigma {\rm Tr} ({\bf m}^* U e^{i\theta/N_f} + {\bf m}
U^{-1} e^{-i\theta/N_f})],
\label{eff2}
\ee
where we have split the integration over $U$ in a $U(1)$ integral over $\theta$
and an integral over $SU(N_f)$ $(U\rightarrow U \exp[i\theta/N_f])$.
For a diagonal mass matrix the condensate is given by
\be
\langle \bar q_f q_f \rangle = \frac 1{2n} \del_{m_f} \log Z,
\label{condens}
\ee
where the differentiation is with respect to $one$ of the quark masses.
The right hand side should be evaluated for $m_fn\Sigma \gg 1$, which allows
us to use a saddle point integration for $U$ with the result that
\be
\langle \bar q_f q_f \rangle = \Sigma \cos (\frac{\theta}{N_f}).
\ee
This completes the calculation of the partition function that, with
the identification of $2n$ as the volume of space-time, was
the starting point of ref. \cite{LS}.

\vskip 1.0 cm
\noindent
\subsection{Effective theory for $\beta = 1$}
\vskip 0.5 cm
For $\beta =1$, the overlap matrix $T$ is real. After averaging over $T$ the
partition function is given by
\be
Z_1(\bar\nu, N_f) \sim
\int {\cal D} \psi^* {\cal D} \psi &&\exp\left [ \right .
-\frac 1{2 n\Sigma^2 \beta}\left [
\left ( \begin{array}{c} \psi_{R\,i} \\ \psi^{*}_{R\,i} \end{array}\right)_f
I_{ff'}
\left ( \begin{array}{c} \psi_{L\,k} \\ \psi^{*}_{L\,k}\end{array}\right)_{f'}
\right ]^2 \nonumber \\ &+&
{\bf m}^*_{fg}\psi^{f\,*}_{R\, i}
\psi^{g}_{R\, i}+{\bf m}_{fg}\psi^{f\,*}_{L\, k}\psi^{g}_{L\, k}
\left . \right ].
\ee
The introduction of anti-symmetric unit matrix
\be
 I= \left ( \begin{array}{cc} 0 & -{\bf 1}\\ {\bf 1} & 0 \end{array}\right ),
\ee
allows us to rearrange the fermions in multiplets of length $2N_f$,
in agreement
with the well-known result \cite{Shifman-three} that for two colors the
the chiral symmetry group is enlarged from $U(N_f)\times U(N_f)$ to
$U(2N_f)$.
Since baryons consist of two quarks for $N_c = 2$, it is
not surprising that the four-fermion interaction contains
both mesonic and di-quark bilinears.
It can be rewritten as the difference of
\be
\frac 14\left [\left ( \begin{array}{c} \psi_{R\,i} \\ \psi^{*}_{R\,i}
\end{array}\right)_f
\left ( \begin{array}{c} \psi_{R\,i} \\ \psi^{*}_{R\,i}
\end{array}\right)_g+
I_{gg'} \left ( \begin{array}{c} \psi_{L\,k} \\ \psi^{*}_{L\,k}
\end{array}\right)_{g'}
\left ( \begin{array}{c} \psi_{L\,k} \\ \psi^{*}_{L\,k}
\end{array}\right)_{f'} I_{f'f}\right ]^2
\ee
and a similar expression with the plus sign exchanged by a minus sign.
Each of the fermionic bilinears is anti-symmetric in the flavor indices.
Using the Hubbard-Stratonovitch transformation (5.3),
the squares can be linearized with the help of an anti-symmetric matrix
which yields the following partition function
\be
Z_1(\bar\nu, N_f) \sim
\int {\cal D} \psi^* {\cal D} \psi \exp\left [ \right .
-2n\Sigma^2 \beta{\rm Tr}(AA^\dagger)
&+& \left (\begin{array}{c} \psi_R \\ \psi_R^*
\end{array} \right )_f
\left ( A^\dagger + \frac 12 {\cal M}^*\right)_{fg}
\left (\begin{array}{c}\psi_R \\ \psi_R^* \end{array}\right)_g
\nonumber \\
&+&\left (\begin{array}{c}\psi_L\\ \psi_L^* \end{array}\right)_f
\left (I A I^T + \frac 12{\cal M}\right )_{fg}
\left (\begin{array}{c} \psi_L \\ \psi_L^* \end{array}\right)_g\left .\right ].
\nonumber \\
\label{realall}
\ee
The mass matrix is defined by
\be
{\cal M} = \left ( \begin{array}{cc} 0 & -{\bf m}\\ {\bf m} & 0
\end{array} \right ),
\ee
and $A$ is a general antisymmetric complex $2N_f \times 2 N_f$
matrix. Carrying out the fermion integrals and rescaling $A$ by a factor two,
the partition function can be compactly written as
\be
Z_1(\bar \nu, N_f)
\sim \int {\cal D} A {\rm Pf}^{n+\nu}( A^\dagger+{\cal M}^*) {\rm Pf}^n
(A+{\cal M}) \exp[-\frac{n\beta \Sigma^2}2
{\rm Tr} A A^\dagger],
\label{apart}
\ee
where Pf denotes the Pfaffian of the matrix (For an even dimensional
anti-symmetric matrix,
the Pfaffian is equal to the square root of its determinant with a definite
choice for its phase \cite{Mehta-red,SV}).
The  complex antisymmetric matrix $A$ can be brought into a standard form as
\be
A = U \Lambda U^T,
\label{aulamu}
\ee
where $U$ is a unitary matrix  and $\Lambda$ a real  antisymmetric matrix
with $\Lambda_{k, k+1} = - \Lambda_{k+1, k} =
\lambda_k, k = 1,\cdots, 2N_f -1$ and
all other matrix elements zero. By redefining $U$ we can always choose
all $\lambda_k\ge 0$. In (\ref{apart})
we use $\Lambda$ and $U$ as new
integration variables. Note that $U\in U(2N_f)/(Sp(2))^{N_f}$ so that the
total number of degrees of freedom on both sides of (\ref{aulamu}) is the same.
We are interested in the limit $n\rightarrow\infty$ and
$\|{\cal M}\|\Sigma \ll 1$.
Then the $\Lambda$ integration
can be performed by a saddle point integration at ${\cal M} = 0$, whereas the
remaining integrals have to be performed exactly for the actual value of the
mass. At the saddle point inside
the integration manifold, i.e.
$\lambda_k = 1/\Sigma$, the integrand
does not depend on $U$ for ${\cal M}=0$. For ${\cal M}\ne 0$
the $U$ dependence is in the form
$U I U^T$, where $I$ is the antisymmetric unit matrix.
The integration over the soft modes $U$ is thus
parametrized by the coset $U(2N_f)/Sp(2N_f)$. For $\|{\cal M}\|\Sigma \ll 1$
the exponentiated determinants can be expanded to first order
in ${\cal M}$ resulting in the low energy finite volume partition function
\be
Z_1(\bar\nu, N_f) \sim \int d\theta \exp(i\bar\nu \theta) \int_{U \in
SU(2N_f)/Sp(2N_f)} \exp [n \Sigma {\rm Re}(e^{i\theta/N_f}
{\rm Tr} U I U^T {\cal M}) ],
\label{eff1}
\ee
which was used as starting point for the calculation of the Leutwyler-Smilga
sum rules in \cite{SV} (with the identification of $2n$ as the volume of
space-time). The phase of $\det U$ has been isolated by
the substitution $U \rightarrow U \exp[ i\theta/{2N_f}]$. Note that in this
way the phase of the Pfaffian in (5.16) covers the full complex unit circle for
$\theta \in [0,2\pi]$. The integration
over the stability subgroup only modifies the partition function by a
constant. Therefore, the integration in (\ref{eff1}) can be extended
to $SU(N_f)$, which facilitates further evaluation of the partition
function \cite{SV}.

Also in this case the condensate is given by the logarithmic derivative in
(\ref{condens}) evaluated for $m_fn\Sigma \gg 1$ (with $m_f$ an eigenvalue of
${\bf m}$). In this limit the $U$
integral can be performed by a saddle point method. The saddle point is at
$UIU^T = 1$ resulting in the identification
\be
\langle \bar q_f q_f \rangle = \Sigma \cos(\frac{\theta}{N_f}).
\ee

\vskip 1.0 cm
\subsection{Effective theory for $\beta = 4$}
\vskip 0.5 cm
In this case the overlap matrix elements are quaternion real
\be
T_{ik} = \sum a^\mu_{ik} i\sigma^+_\mu,
\ee
where $\sigma^+_\mu = (-i, \vec \sigma)_\mu$ and the $a^\mu_{ik}$ are real.
In terms of the $a_\mu$ variables
the partition function reads
\be
Z_4(\bar\nu, N_f) \sim \int {\cal D} a^\mu {\cal D}\psi^* {\cal D} \psi
&&\exp\left [\right . - \frac{n\beta \Sigma^2}2 {\rm Tr} a_\mu a_\mu^T+
\psi^{f\,*}_{R\, i} a^\mu_{ik} \sigma_\mu^+ \psi_{L\,k}^f
-\psi^{f\,*}_{L\, k} a^\mu_{ik} (\sigma_\mu^+)^\dagger \psi_{R\,i}^f
\nonumber\\&+&
{\bf m}^*_{fg}\psi^{f\,*}_{R\, i}
\psi^{g}_{R\, i}+{\bf m}_{fg}\psi^{f\,*}_{L\, k}\psi^{g}_{L\, k}\left .\right
].
\label{z4}
\ee
The components of the fermionic variables are 2 component vectors defined by
\be
\psi_R = \left ( \begin{array}{c} \chi_R \\ \chi_R^* \end{array} \right ),
\qquad
\psi_L = \left ( \begin{array}{c} \chi_L \\ \chi_L^* \end{array} \right ).
\label{weyl}
\ee
By using the identity $((\sigma_\mu^+)^\dagger)^T = - \sigma_2 \sigma_\mu^+
\sigma_2$ we find that the two fermionic terms in (\ref{z4}) are
identical. Averaging over the overlap matrix elements leads to
\be
Z_4(\bar\nu, N_f) \sim \int {\cal D}\psi^* {\cal D} \psi
\exp[\frac 2{n\beta\Sigma^2} \psi^{f\,*}_{R\, i}  \sigma_\mu^+ \psi_{L\,k}^f
\psi^{g\,*}_{R\, i} \sigma_\mu^+ \psi_{L\,k}^g
+ {\bf m}^*_{fg}\psi^{f\,*}_{R\, i}
\psi^{g}_{R\, i}+{\bf m}_{fg}\psi^{f\,*}_{L\, k}\psi^{g}_{L\, k}].
\nonumber\\
\ee
Using the Fierz identity
\be
\sum _\mu \sigma_\mu^{+\, \alpha \beta} \sigma_\mu^{+\, \gamma \delta}
= 2(\delta_{\alpha \delta} \delta_{\gamma\beta} - \delta_{\alpha \beta}
\delta_{\gamma \delta})
\ee
and the representation (\ref{weyl}) of the spinors,
the partition function can be simplified to
\be
Z_4(\bar\nu, N_f) \sim \int {\cal D}\psi^* {\cal D} \psi
&&\exp[ \frac 4{n\beta\Sigma^2}(\chi^{f\,*}_{R\, i}  \chi_{R\,i}^g
+\chi^{g\,*}_{R\, i} \chi_{R\,i}^f) (\chi^{f\,*}_{L\, k}  \chi_{L\,k}^g
+\chi^{g\,*}_{L\, k} \chi_{L\,k}^f) \nonumber \\
&+&{\bf m}^*_{fg}\psi^{f\,*}_{R\, i}
\psi^{g}_{R\, i}+{\bf m}_{fg}\psi^{f\,*}_{L\, k}\psi^{g}_{L\, k}].
\nonumber\\
\ee
This four fermion interaction can be written as the difference of two squares
which can be linearized with the help of the Hubbard-Stratonovitch
transformation (\ref{Hubbard})
by introducing the bosonic variables $\sigma_1$ and
$\sigma_2$. Both $\sigma_1$ and $\sigma_2$ are symmetric real valued
$N_f \times N_f$ matrices. In terms of the new variables
the partition function reads
\be
Z_4(\bar\nu, N_f) &\sim& \int {\cal D}\psi^* {\cal D} \psi {\cal D} \sigma_1
{\cal D} \sigma_2 \exp [
-\frac {n\beta\Sigma^2}4 {\rm Tr }(\sigma_1^2 + \sigma_2^2)] \nonumber \\
&\times& \exp[ 2\chi^{f\,*}_{R\, i}  \chi_{R\,i}^g
(\sigma_1 -i\sigma_2)_{fg}+2\chi^{f\,*}_{L\, k}  \chi_{L\,k}^g
(\sigma_1 +i\sigma_2)_{fg}
+2({\bf m}^*_{fg}\chi^{f\,*}_{R\, i}  \chi_{R\,i}^g +{\bf m}_{fg}
\chi^{f\,*}_{L\, k}  \chi_{L\,k}^g)].
\nonumber\\
\ee
The Grassmann integrals can be performed trivially resulting in the
partition function
\be
Z_4(\bar\nu, N_f) \sim \int {\cal D}A {\det}^{n+ \nu} (A^\dagger+ {\bf m }^*)
{\det}^n (A+ {\bf m}) \exp[ - \frac {n\beta \Sigma^2}4 {\rm Tr}A A^\dagger],
\label{za4}
\ee
where $A$ is a complex symmetric matrix. Such a real matrix can be diagonalized
by a unitary matrix as follows
\be
A = U \Lambda U^T
\ee
with $\Lambda$ a real nonnegative diagonal matrix.
As before we use $\Lambda$ and $U$
as new integration variables. The mass plays the role of a small symmetry
breaking term ($\|{\bf m}\|\Sigma \ll 1$)
so that the integration over the hard modes,
the $\Lambda$,
can be performed by a saddle point approximation at ${\bf m} = 0$. The saddle
points are given by
\be
\lambda_k = \pm \frac 1{\Sigma},
\ee
but only the saddle point with all signs positive is inside the integration
manifold. The integral over the soft modes has to be taken into account
exactly, but the exponentiated determinant can be expanded to first order in
${\bf m}$. Writing $U$ as the product of a phase and a special unitary
transformation $(U \rightarrow \exp(i\theta/2N_f) U)$
we arrive at the partition function
\be
Z_4(\bar\nu, N_f) \sim \int d\theta e^{i\bar\nu\theta}\int_{U \in
SU(N_f)/O(N_f)}
{\cal D} U \exp[2n\Sigma {\rm Re} ( e^{i\theta/N_f}{\rm Tr} U U^T {\bf m})].
\label{eff4}
\ee
Because the integrand depends only on $U^T U$, the integration over
$U$ is effectively over the coset $SU(N_f)/O(N_f)$, where the volume of
the stability subgroup can be absorbed in the normalization.
As before, the condensate is given by the logarithmic derivative of the
partition function for $m_f n\Sigma \gg 1$ (The eigenvalues of
${\bf m}$ are denoted by $m_f$). The integral can than be performed
by a saddle point method resulting in the identification
\be
\langle \bar q_f q_f \rangle = \Sigma\cos(\frac{\theta}{N_f}).
\ee
With the identification of $2n$ as the volume of space-time,
the partition function (\ref{eff4})
was the starting point for the derivation of the Leutwyler-Smilga
sum rules in \cite{LS,SV}.

\vskip 1.5 cm
\renewcommand{\theequation}{6.\arabic{equation}}
\setcounter{equation}{0}
\section{Sum rules for the eigenvalues of the Dirac operator}
\vskip 0.5 cm

As already mentioned in the introduction, the sum of inverse powers of
the eigenvalues of the Dirac operator satisfies sum rules. Since they
have been derived both from effective field theory and random matrix theory,
we want to restrict ourselves to
one feature that was not clarified in earlier work.

To obtain sum rules we expand both the expectation value of the fermion
determinant and the finite volume partition function in powers of $m^2$
(We consider the case of a diagonal mass matrix with all masses equal to $m$).
For the fermion determinant we obtain
\be
\frac{Z(m)}{Z(0)} = 1 + m^2 N_f \sum_{\lambda_n > 0} \frac 1{\lambda_n^2},
\label{detmm}
\ee
where, for simplicity, we consider the case with all masses equal.
Equating the coefficients of $m^2$ in (6.1) and of the
expansion of the finite volume partition function (see below), leads to
sum rules that can be summarized into one formula \cite{SV}
valid for each of the three cases discussed in section 5:
\be
\frac 1{V^2}\sum_{\lambda_n > 0} \frac 1{\lambda_n^2} =
\frac{\Sigma^2}{4(|\nu|+({\rm dim}({\rm coset})+1) /N_f)}.
\label{sumrule}
\ee
The volume of space-time is denoted by $V$.
In this section we give a derivation for $\nu = 0$
showing that this unifying feature was no accident.

The coefficient of $m^2$ of the ratio $Z(m)/Z(0)$ for $\nu = 0$ of finite
volume partition functions
given in (\ref{eff2}), (\ref{eff1}) and (\ref{eff4}) involves the calculation
of integrals of the form
\be
\zeta(A)=\int_{V \in G/H} dV {\rm Tr}( V A){\rm Tr}^*( V A),
\label{inteff}
\ee
For $\nu = 0$ the $U(1)$ integral over $\theta$ can be absorbed in the
integration over the coset $G/H$, which now becomes
$U(2N_f)/Sp(2N_f)$, $U(N_f)$ and  $U(N_f)/SO(N_f)$
for $\beta = 1$, 2 and 4, respectively.
The matrix $A$ is the mass matrix which
is real anti-symmetric, complex  and real symmetric in this order.
Using the invariance of the measure it follows that in
all three cases $\zeta(A) \sim {\rm Tr } A A^\dagger$.
For a mass matrix
with equal masses in the standard form,
$A\sim {\bf 1} $ for $\beta =2$ and
$\beta =4$  and proportional to the antisymmetric
unit matrix $I$ for $\beta = 1$.

To proceed let us introduce generators $t_k$ of the cosets
\cite{SV}. They satisfy the
orthogonality relations
\be
{\rm Tr}\, t_k t_l = \frac 12 \delta_{kl},
\label{ortho}
\ee
and are real anti-symmetric, Hermitean and real symmetric, for
$\beta=1$, $\beta=2$  and $\beta = 4$, respectively.
The total number of generators is denoted by $M$ and the dimension of the
matrices $U$ and $t_k$ is denoted by $d$.
The generators will be chosen such that
$t_1$ is proportional to the mass matrix. Because of the
invariance discussed in the previous paragraph and the normalization
(\ref{ortho}), we find that
\be
\zeta(t_1) = \zeta(t_2) = \cdots = \zeta(t_M),
\ee
allowing us to rewrite the integral as
\be
\zeta(t_1) = \frac 1M \sum_{k=1}^M\int_{V \in G/H} dV
{\rm Tr}( V t_k){\rm Tr}^*(V  t_k).
\ee
By expanding $U$ in its generators ($U = \sum_k u_k t_k$) and using the
orthogonality relation (\ref{ortho}) and the unitarity of $U$, the integral
becomes trivial. The result is given by
\be
\zeta(t_1)=\frac d{2M} {\rm vol}(G/H).
\ee
The volume of the coset cancels in the ratio $Z(m)/Z(0)$ of the partition
functions. Using that
the dimensionality $d = N_f$ for $\beta =2$ and $\beta =4$ and $d= 2N_f$
for $\beta = 1$, we find that
\be
\left .\frac{Z(m)}{Z(0)} \right |_{\nu = 0}=
1 + \frac 14 m^2 V^2 \Sigma^2 \frac {N_f}{2M} 2N_f.
\label{effmm}
\ee
The extra factor $2N_f$ appeared because of the difference in normalization
between the unit matrix and the generators (\ref{ortho}). As already discussed
in section 4, we have identified the total number of zero modes, $2n$,
with the volume $V$ of space-time.

In (\ref{sumrule}) the dimension of the coset does not include the $U(1)$
factor, so in the notation of (\ref{sumrule})  we have
\be
M = {\rm dim(coset)} + 1.
\ee
By comparing the coefficients of $m^2$ in (\ref{detmm}) and (\ref{effmm})
we indeed reproduce the sum rule (\ref{sumrule}).

\vskip 1.5 cm
\section{Conclusions}
\vskip 0.5 cm

Starting from a chiral random matrix theory with the symmetries of the Dirac
operator in an arbitrary background field, we have derived
the finite volume partition functions that were used as starting point in the
derivation of the Leutwyler-Smilga sum rules in \cite{LS,SV}.
A derivation of the simplest Leutwyler-Smilga sum rules
shows  in a natural way that they depend on the number of Goldstone
bosons per flavor.

Our results explain the miracle that
effective Lagrangians and chiral random matrix
theory produce the same sum rules. The advantage of chiral random matrix
theory is that it leads naturally to three classes of sum rules. The absence
of a third class in \cite{LS} motivated the studies in \cite{V} with the
result that QCD with two colors in the fundamental representation constitutes
a separate universality class. In the present work we have shown that then
the Dirac operator is real and that the corresponding random matrix
theory (chGOE) is equivalent to an effective theory based on the coset
$SU(2N_f)/Sp(2N_f)$. For three or more colors in the fundamental representation
the Dirac operator is complex and the corresponding random matrix theory
with complex matrix elements (chGUE) is equivalent to a finite volume partition
function based on the coset $SU_R(N_f)\times SU_L(N_f) /SU(N_f)$. The last
universality class is for a Dirac operator with adjoint fermions and two or
more colors. Then the Dirac operator is quaternion real, and in this paper
we have shown that the corresponding random matrix theory with quaternion
real matrix elements is equivalent to a finite volume partition function based
on the coset $SU(N_f)/O(N_f)$.

Apart from two external parameters, the vacuum angle and the mass matrix,
the random matrix theory and the corresponding
finite volume partition function  depend on only one dynamical parameter
of QCD: the chiral condensate. The finite volume partition function, and thus
the chiral random matrix theory, provides
us with the mass dependence of the QCD partition function.
However, this does not imply that all properties of the chiral random
matrix model are physical. For example, the average level
density, which has a semicircular shape, is certainly unphysical. This leads to
the question which properties
of the chiral random matrix theory actually determine the finite volume
partition function.

In the finite volume partition function the mass occurs in the combination
$m V$, whereas in the QCD partition function or in the random matrix theory,
the mass occurs as $m/\lambda$, where $\lambda$ is an eigenvalue of the
Dirac operator. This suggests that the effective theory is only sensitive
to the distribution of eigenvalues on a scale $1/V$. The spectral density
on this scale, also called the microscopic spectral density, is well defined.
{}From the work on quantum
chaos and mesoscopic systems (including nuclei) \cite{univers}
we know that the distribution
of eigenvalues on a microscopic scale is universal and is given by random
matrix theory. The most impressive piece of work in the present context is
by Slevin and Nagao \cite{SLEVIN-NAGAO} who
studied the logarithm of the transfer
matrix of a mesoscopic system in a magnetic field for the
Hofstadter model. The random matrix theory of this
system has the symmetries of the Dirac operator for $N_f = 0$ and three
or more colors in the fundamental representation. Indeed, by numerical
computations, they found that the microscopic spectral density
is given by the corresponding  chiral random matrix theory.
This led us to the conjecture
that the so called microscopic spectral density is universal. However, this
does not imply that it can be determined from the finite volume
partition function.
For example, for one flavor, in each of the
three cases, Goldstone bosons are absent and
the finite volume partition function is the same
but the random matrix theory is different. This shows that
the low-lying spectrum of the Dirac operator cannot be derived from the
complete
set of Leutwyler-Smilga sum rules.

We conclude with the statement that although the finite volume partition
function does not determine the microscopic spectral density, there
is ample evidence that if combined with universality according to
the anti-unitary symmetries of the Dirac operator this leads to a unique
prediction of the spectral density of the Dirac operator on a scale of
no more than a finite number of eigenvalues from zero.

\vskip 1.5 cm
\vglue 0.6cm
{\bf \noindent  Acknowledgements \hfil}
\vglue 0.4cm
 The reported work was partially supported by the US DOE grant
DE-FG-88ER40388.

\vfill
\eject
\newpage

\end{document}